\documentclass[prd,showkeys,floatfix,nofootinbib,
               preprint,12pt,
               tightenlines,fleqn]{revtex4} 

\usepackage{amsmath,amssymb,revsymb,graphicx,dcolumn}
\usepackage{leftidx}

\newcommand{\beq}{\begin{equation}}
\newcommand{\eeq}{\end{equation}}
\newcommand{\beqa}{\begin{eqnarray}}
\newcommand{\eeqa}{\end{eqnarray}}
\newcommand{\bsubeqs}{\begin{subequations}}
\newcommand{\esubeqs}{\end{subequations}}

\newcommand{\vin}{\rotatebox[origin=c]{-90}{$\in$}}

\usepackage{enumerate}

\begin{document}
\title[]
      {Can gravitational collapse and black-hole \\ evaporation
be a unitary process after all?\vspace*{5mm}}
\author{Slava Emelyanov}
\email{viacheslav.emelyanov@physik.uni-muenchen.de}
\affiliation{Arnold Sommerfeld Center for
Theoretical Physics,\\
Ludwig Maximilian University (LMU),\\
80333 Munich, Germany\\}

\begin{abstract}
\vspace*{2.5mm}\noindent
This paper shows a way of how to resolve the non-unitarity problem in black-hole physics without 
modifications of the basic principles of local quantum field theory.
\end{abstract}


\date{\today}

\maketitle

\section{Introduction}

Right after Hawking's discovery of the black-hole evaporation~\cite{Hawking1}, it was realized by him~\cite{Hawking}
(see also~\cite{Page,Giddings1,Wald1,Carlip} and~\cite{Frolov&Novikov,Fabbri&Navarro-Salas,Frolov&Zelnikov}) 
that this effect leads to a violation of one of the basic principles of local quantum 
field theory (LQFT), namely the unitarity. This problem is also known as the incompatibility of the quantum theory principles with the general relativity ones
or the breakdown of \emph{quantum} predictability.\footnote{The breakdown of \emph{classical} predictability 
is related with the singularity inside the black hole, see, e.g.~\cite{Fabbri&Navarro-Salas}.} 

Nowdays it seems one awaits that the non-unitarity can be resolved at the level of quantum gravity. In this respect,
the most promising role is expected to be played by the anti-de Sitter/conformal field theory (AdS/CFT) correspondence (for instance,
see~\cite{Hawking2}).

One more possible resolution could be through rejecting some of the LQFT principles in favour of the unitarity. For example,
this may be locality~\cite{Giddings} (and references therein).

We will show in this paper how one may resolve the problem without modifications of the LQFT basic principles at the semi-classical
approximation. In other words, we will show how classical theory of gravity, i.e. general relativity, and LQFT of
matter fields could be still compatible with each other during gravitational collapse forming a black hole and its
subsequent evaporation.

We will work within the algebraic framework of local quantum field theory. Specifically, we consider a scalar, non-interacting 
field $\hat{\Phi}(x)$ in the spacetime geometry denoted as $\mathcal{M}$ below. This field operator can by used to
generate the so-called operator algebra $\mathcal{A}(\mathcal{M})$. This algebra is composed of an identity operator 
$\hat{\mathbf{1}}$ and finite sums of finite
products of the field $\hat{\Phi}(x)$ smeared out over test functions $\{f(x),\;x \in \mathcal{M}\}$, i.e. functions being smooth and compactly supported.
States are defined as linear, 
positive and normalized functionals on $\mathcal{A}(\mathcal{M})$. One can associate a certain Hilbert
space representation $\mathcal{H}$ of the algebra with a given state (through the Gelfand-Naimark-Segal construction). 
For more on this, see a basic reference~\cite{Haag} and recent 
reviews~\cite{Hollands&Wald,Khavkine&Moretti,Fewster&Verch} devoted to LQFT in curved classical spacetimes.

Our main strategy is to extensively exploit the LQFT principles, properties and their subtle mathematical consequences, namely
\begin{enumerate}[(a)]

\item there exist unitarily inequivalent Fock space representations of the same canonical commutation 
relation in quantum field theory;

\item a primary interpretation of quantum field theory is given in terms of local operations;

\item a factorization of a field operator algebra $\mathcal{A}(\mathcal{M})$ into a product of two commuting subalgebras 
$\mathcal{A}_1(\mathcal{M}_1){\otimes}\mathcal{A}_2(\mathcal{M}_1)$, where $\mathcal{M}_1{\cup}\mathcal{M}_2 \subset \mathcal{M}$
and the Cauchy surface $\Sigma_\mathcal{M} = \Sigma_{\mathcal{M}_1}{\cup}\Sigma_{\mathcal{M}_2}$, 
does not lead to a Hilbert space factorization.
\end{enumerate} 
The explanation of these items in quantum field theory are in order: (a) The existence of unitarily inequivalent Fock space representations 
of $\mathcal{A}(\mathcal{M})$ is related to the non-separability of 
its Hilbert space representation~\cite{Haag, Umezawa,Blasone&Jizba&Vitiello}; 
(b) Since spacetime isometry plays a crucial role in defining field excitations, an interpretation of quantum field theory is generally impossible in terms of excitations~\cite{Birrell&Davies}. However, quantum field theory can be always interpreted in terms of local 
operations~\cite{Haag}; (c) This is a subtle mathematical fact which will be exemplified below.

The outline of this paper is as follows. In Sec.~\ref{sec:adsbh}, we will show how the gravitational collapse can be still a unitary process. 
In Sec.~\ref{sec:conclusions}, we will reinterpret the black-hole evaporation, which does not lead then 
to the information loss problem.

Throughout this paper the fundamental constants are set to unity, $c = G = k_\text{B} = \hbar = 1$.

\section{The proposal}
\label{sec:adsbh}

We will follow the notations of reference~\cite{Hawking} in order to simplify our consideration
and to avoid possible misunderstandings of the key point.

\subsection{Operator algebra $\mathcal{A}(\mathcal{M})$ and its representations}
\label{subsec:algebra}

One of the advantages of working within the algebraic framework is that one does not need at the outset 
to choose a concrete Hilbert space representation of the canonical commutation relation. Instead of
the canonical commutation relation, one can consider a commutator between the quantum field $\hat{\Phi}(x)$ 
at two spacetime points, namely
\beqa\label{eq:as}
\big[\hat{\Phi}(x), \hat{\Phi}(y)\big] &=& i\Delta(x,y)\,,
\eeqa
where $\Delta(x,x')$ is the so-called causal propagator. It is given by a difference between the retarded 
and advanced Green function of the scalar field equation: $\Box\Phi(x) = 0$. One may call $\Delta(x,x')$ as an algebraic
structure in the set $\mathcal{A}(\mathcal{M})$. The spacetime metric of $\mathcal{M}$ is now fixed 
and corresponds to the collapsing matter shell forming a spherical black hole which then evaporates in particular due to the 
backreaction produced by the field $\hat{\Phi}(x)$.

We denote a set of the rest matter-field operators as $\mathcal{A}_\text{m}(\mathcal{M})$. The collapsing 
shell is supposed to be composed of these fields. We denote a physical Hilbert
space representation\footnote{It should be noted that this is actually 
the Fock space representation, i.e. a separable
subspace in the whole non-separable representation of the field operators. It became common 
to call it as a Hilbert space.} 
of $\mathcal{A}(\mathcal{M})$ and $\mathcal{A}_\text{m}(\mathcal{M})$ as $\mathcal{H}$.
The representation $\mathcal{H}$ is built on a physical state 
$|\Omega\rangle$, such that $\mathcal{A}(\mathcal{M}){\otimes}\mathcal{A}_\text{m}(\mathcal{M})|\Omega\rangle$ 
is dense in $\mathcal{H}$. We also introduce $\mathcal{H}_{\Phi} \subset \mathcal{H}$ which is a Hilbert space
generated by operators from $\mathcal{A}(\mathcal{M})$ by acting on $|\Omega\rangle$.

\subparagraph{Shell state.}

If the quantum field under consideration does not contribute to 
an almost local 
operator\footnote{For a definition of an almost local operator, see~\cite{Haag}, Sec. II.4.1.} 
$\hat{\mathcal{O}}_\text{Shell}$ composed of the matter fields building the collapsing shell, 
i.e. $|\Omega_\text{Shell}\rangle = \hat{\mathcal{O}}_\text{Shell}|\Omega\rangle$, then the quantum field $\hat{\Phi}(x)$ is oblivious 
to the shell, but not to the spacetime geometry. Outside of the matter shell, $\hat{\mathcal{O}}_\text{Shell}$ vanishes,
so that the shell state $|\Omega_\text{Shell}\rangle$ becomes $|\Omega\rangle$.
Thus, the state $|\Omega_\text{Shell}\rangle$ is a localized state which looks empty outside of the
support of the operator $\hat{\mathcal{O}}_\text{Shell}$. 

We actually make an assumption
that one can associate to each single particle, say,
a proton, state $|p\rangle$ a localized operator $\hat{\mathcal{O}}_p$ from 
$\mathcal{A}_\text{m}(\mathcal{M}$), 
which is in turn composed of the quark, gluon, electromagnetic field operators. 
This operator generates the state from the vacuum $|\Omega\rangle$.
This should be understood in the weak sense, i.e. $\langle p|\hat{\mathcal{O}}_p|\Omega\rangle \neq 0$. 
The matter shell is then a complicated combinations of such operators of each particle of the
shell. An association of a localized operator with a particle state is actually not a novel idea~\cite{Haag}. It is employed to describe
composite particles (e.g. hadrons) to which one cannot ascribe any fundamental field operator.

The quantum field $\hat{\Phi}(x)$ plays a role of a trial quantum field considered in the gravitational collapse. To make our set-up more realistic, we assume that $\hat{\mathcal{O}}_\text{Shell} \in \mathcal{A}(\mathcal{M}){\otimes}\mathcal{A}_\text{m}(\mathcal{M})$. Note that $|\Omega_\text{Shell}\rangle \in \mathcal{H}$. 

\subparagraph{Black-hole state.}

It may correspond to ranges of Unruh vacua when the evaporating hole is present. We will denote this set of vacua as 
$|\Omega_\text{BH}\rangle$, such that  $|\Omega_\text{BH}\rangle = \hat{\mathcal{O}}_\text{BH}|\Omega\rangle$. The 
almost local 
operator $\hat{\mathcal{O}}_\text{BH}$
with a support of non-vanishing measure 
corresponds to $\hat{\mathcal{O}}_\text{Shell}$ after the black-hole 
formation. 
The evolution of the geometry leads to a change of the algebraic structure of $\mathcal{A}(\mathcal{M})$ as well as
$\mathcal{A}_\text{m}(\mathcal{M})$. However, the composition of $\hat{\mathcal{O}}_\text{BH}$ and 
$\hat{\mathcal{O}}_\text{Shell}$ does \emph{not} vary.

The choice of the Unruh vacua is based on the fact that the quantum field being in 
$|\Omega_\text{BH}\rangle$ does not lead to a divergence of the field stress tensor on the future event horizon. This condition
defines these states (see~\cite{Frolov&Novikov} and~\cite{Dimock&Kay,Dappiaggi&Moretti&Pinamonti} for a more rigorous definition). 
This choice of $|\Omega_\text{BH}\rangle$ allows to have a self-consistent 
consideration of quantum gravity in the semi-classical approximation. The Unruh vacuum at a given moment of time 
will be also called as {\sl system's vacuum}, where the system is composed of the quantum matter and gravity fields.
Note that $|\Omega_\text{BH}\rangle \in \mathcal{H}$.

Up to now there does not exist a widely accepted theory of quantum gravity. Therefore,
in this paper, we employ both
general relativity (GR) and local quantum field theory up to regimes when one cannot trust anymore in 
their predictions. At the level of GR, the collapse of matter is described by a change of the spacetime structure.
Its change leads to an appearance of a singularity at $r = 0$. In the vicinity of this point, GR is certainly unreliable. 
At the level of LQFT, the collapse of matter is described by a unitary evolution of the matter state from
$|\Omega_\text{Shell}\rangle$ to $|\Omega_\text{BH}\rangle$. The black-hole state $|\Omega_\text{BH}\rangle$ 
as noted above has a non-vanishing support in the vicinity of the singularity $r = 0$, e.g. due to the Heisenberg principle. In a sense, there should be a black-hole \emph{core} of the original matter of the collapsing shell 
inside the horizon $r = 2M$. The core is expected to be of a macroscopic size which is dynamically generated.
It is worth noting that it is legitimate to use GR only outside of the support of the 
black-hole core. Whenever one wants to "touch" the state $|\Omega_\text{BH}\rangle$ inside of its support, 
one generally should expect a discovery of quantum gravity degrees of freedom.\footnote{This is analogous to discovering gluons and quarks when one probes the structure of a proton in the high-energy collisions.}
Quantum gravity is expected to resolve the problem of the breakdown of \emph{classical} predictability.

A similar (only to a certain extent) idea of representing a black hole was proposed in~\cite{Dvali&Gomez} and
further investigated in~\cite{Hofmann&Rug,Gruending&Hofmann&Mueller&Rug}.
\vspace*{1.5mm}\noindent

One may also introduce three extra Hilbert spaces $\mathcal{H}_-$, $\mathcal{H}_\text{H}$ and 
$\mathcal{H}_\text{I}$. These are associated with the states $|0_-\rangle$, $|0_\text{H}\rangle$ and 
$|0_\text{I}\rangle$, respectively,~\cite{Hawking}. 

The state $|0_-\rangle$ is {\sl observer's vacuum} (at past time infinity) or the initial vacuum state for scalar particles,  such that
\beqa\label{eq:phi:a}
\hat{\Phi}(x) &=& \hat{a}(x) + \hat{a}^\dagger(x)\,,
\eeqa
where 
\beqa
\hat{a}(x) &=& \sum_i\;f_i(x)\,\hat{a}_i\,,
\eeqa
$\hat{a}(x)|0_-\rangle = 0$ and $f_i(x)$ are mode functions. Following~\cite{Hawking}, we assume that at past 
time infinity observer's vacuum coincides with system's vacuum, i.e. $|0_-\rangle \cong |\Omega_\text{Shell}\rangle$ 
or, in terms of Hilbert spaces, $\mathcal{H}_- \cong \mathcal{H}_{\Phi}$.

The state $|0_\text{I}\rangle$ is {\sl observer's vacuum}  (at future time infinity) or the vacuum state for outgoing scalar particles. The state
$|0_\text{H}\rangle$ is the vacuum state for particles falling into the black hole. One can equivalently rewrite \eqref{eq:phi:a}
as an operator equality as follows
\beqa\label{eq:phi:bc}
\hat{\Phi}(x) &=& \hat{b}(x) + \hat{b}^\dagger(x) + \hat{c}(x) + \hat{c}^\dagger(x)\,,
\eeqa
where $\hat{b}(x)|0_\text{I}\rangle = 0$ and $\hat{c}(x)|0_\text{H}\rangle = 0$.

Following~\cite{Hawking}, one defines the final scalar particle vacuum state, $|0_+\rangle$, as 
$|0_\text{H}\rangle{\otimes}|0_\text{I}\rangle$. Hence, the Hilbert space representation of the algebraic structure
\eqref{eq:as} corresponds to the factorized product $\mathcal{H}_\text{H}{\otimes}\mathcal{H}_\text{I}$. 

\subsection{Splitting of operator algebra $\mathcal{A}(\mathcal{M})$}
\label{subsec:splitting}

The field operators $\hat{\mathcal{O}}_b(x)$ composed of $\{\hat{b}(x),\hat{b}^\dagger(x)\}$ and $\hat{\mathcal{O}}_c(x)$ 
composed of $\{\hat{c}(x),\hat{c}^\dagger(x)\}$ commute with each other~\cite{Hawking}. This implies that one can express 
the total algebra of the field operators $\mathcal{A}(\mathcal{M})$ as a factorized product of two commuting operator algebras
i.e. $\mathcal{A}(H){\otimes}\mathcal{A}(I)$, such that
\bsubeqs
\beqa
\hat{\mathcal{O}}_a(x) &\in& \mathcal{A}(\mathcal{M})\,,
\\[1mm]
\hat{\mathcal{O}}_b(x) &\in& \mathcal{A}(I)\,,
\\[1mm]
\hat{\mathcal{O}}_c(x) &\in& \mathcal{A}(H)\,,
\eeqa
\esubeqs
where field operators $\hat{\mathcal{O}}_a(x)$ are composed of $\{\hat{a}(x),\hat{a}^\dagger(x)\}$. This splitting of the
total algebra can also be expressed as 
$\hat{\mathcal{O}}_a(x) = \hat{\mathcal{O}}_c(x){\otimes}\hat{\mathcal{O}}_b(x)$.
Note that operators of the form $\hat{\mathcal{O}}_b(x)$ have vanishing support inside of the hole, whereas 
operators of the form $\hat{\mathcal{O}}_c(x)$ outside of the hole.

\subsection{Unitary inequivalence of $\mathcal{H}_{\Phi}$ and $\mathcal{H}_\text{H}{\otimes}\mathcal{H}_\text{I}$}
\label{subsec:inequivalence}

\subparagraph*{Type III property of $\mathcal{A}(\mathcal{N})$.}

There is a subtle mathematical result related to the operator 
factor subalgebras of 
$\mathcal{A}(\mathcal{N})$ in quantum
field theory. This corresponds to 
their  
type III property.
This essentially means 
that a splitting of $\mathcal{A}(\mathcal{N})$ into a factorized product of two its subalgebras 
$\mathcal{A}(\mathcal{N}_1){\otimes}\mathcal{A}(\mathcal{N}_2)$, where $\mathcal{N}_1{\cup}\mathcal{N}_2 \subset \mathcal{N}$
and the Cauchy surface of $\mathcal{N}$ being $\Sigma_{\mathcal{N}} = \Sigma_{\mathcal{N}_1}{\cup}\Sigma_{\mathcal{N}_2}$,
does {\sl not} lead to a Hilbert space factorization, i.e. $\mathcal{H} \not\cong\mathcal{H}_1{\otimes}\mathcal{H}_2$.

Specifically, the factor subalgebra $A(\mathcal{N}_{1})$ is said to be of the type III if for every
projection $\hat{E} \in A(\mathcal{N}_{1})$, there exists an element $\hat{W} \in A(\mathcal{N}_{1})$
that maps $\mathcal{H}$ \emph{isometrically} onto $\hat{E}\mathcal{H}$. In other words, 
$\hat{W}^*\hat{W} = \hat{\bf{1}}$ and $\hat{W}\hat{W}^* = \hat{E}$. This implies that $\hat{E}$ is 
equivalent to $\hat{\bf{1}}$, i.e. trivial. Thus, the subalgebra $A(\mathcal{N}_{1})$ of local operators 
$A(\mathcal{N})$ in
$\mathcal{N}_1$ still acts \emph{irreducibly} on $\mathcal{H}$. 

Note that this is not the case in quantum mechanics, because there one deals with systems with a finite number 
of degrees of freedom. One expresses this as operators in quantum mechanics are of the type I.

This fact was recently emphasized in~\cite{Giddings} with applications to an entanglement. The difference between the type 
I and type III properties of operator algebras are contrasted in~\cite{Yngvason}.

\subparagraph*{Example 1: eternal Schwarzschild geometry.}

Following~\cite{Israel}, one may rewrite the Hartle-Hawking state $|\Omega_\text{HH}\rangle$ as a thermal-field double 
state~\cite{Umezawa}, i.e.
\beqa\label{eq:tfds1}
|\Omega_\text{HH}\rangle &=& \frac{1}{Z^{\frac{1}{2}}}\prod_{\omega lm}\,
\sum\limits_{n = 0}^{+\infty}\,e^{-\beta E_{\omega,n}/2}\,|n_\text{L}\rangle{\otimes}|n_\text{R}\rangle\,,
\quad\text{where} \quad E_{\omega,n} \;\;\equiv\;\; \omega\,n
\eeqa
and $Z$ is a normalization factor. The frequency $\omega > 0$ is defined with respect to the vector 
$\partial_{t_\text{S}}$ ($t_\text{S}$ - the Schwarzschild time coordinate), and $l,m$ are the orbital and magnetic 
numbers referring to a particular representation of the rotational symmetry of the black hole. The inverse temperature
$\beta$ is given by $1/T_\text{H}$, where $T_\text{H}$ is the Hawking (H) temperature~\cite{Hawking1}.
The states entering the right-hand side of \eqref{eq:tfds1} are defined as
\beqa
|n_\text{L}\rangle &\equiv& \frac{1}{\sqrt{n!}}\,
\big(\hat{b}_{\text{L},\omega lm}^{\dagger}\big)^n|\Omega_\text{LB}\rangle\,,
\eeqa
and the same for $|n_\text{R}\rangle$ with $\text{L} \rightarrow \text{R}$ in the above formula.\footnote{Note that
$n_{\text{L},\text{R}}$ depend on $\omega$, $l$ and $m$. Thus, one should not understand $E_{\omega,n}$ as 
the discrete energy levels of the excitations which are actually continuous.} The states $|\Omega_\text{LB}\rangle$ 
and $|\Omega_\text{RB}\rangle$ are the ``left" and ``right" Boulware vacua. One can associate two Hilbert 
spaces, $\mathcal{H}_\text{LB}$ and $\mathcal{H}_\text{RB}$, to both these states. The state on the left-hand side of 
\eqref{eq:tfds1} is the only non-singular state on the black hole horizons.

The normalization factor $Z$ in \eqref{eq:tfds1} is, rigorously speaking, infinite:
\beqa
Z &=& \exp\left(+\frac{\pi}{96M}\,\delta(0)\sum_{l = 0}^{+\infty}(2l+1)\right),
\eeqa
where $M$ is a mass of the black hole.
This can be stressed out by saying 
that the equality \eqref{eq:tfds1} is merely \emph{formal}. It is well-discussed in~\cite{Giddings} (and see also~\cite{Umezawa,Wald}). 
It is worth emphasizing that it is thus {\sl illegitimate} to interpret \eqref{eq:tfds1} as 
$|\Omega_\text{HH}\rangle \in \mathcal{H}_\text{LB}{\otimes}\mathcal{H}_\text{RB}$ in quantum field theory.
It means $\mathcal{H}_\text{HH}$ corresponding to the Hartle-Hawking state $|\Omega_\text{HH}\rangle$ is {\sl not} 
unitarily equivalent to the factorized product $\mathcal{H}_\text{LB}{\otimes}\mathcal{H}_\text{RB}$.

\subparagraph*{Example 2: gravitational collapse.}

Following~\cite{Hawking}, one may rewrite the initial vacuum state $|0_-\rangle$ or $|\Omega_\text{Shell}\rangle$ 
($\cong |\Omega_\text{BH}\rangle$) as a thermal-field double state
\beqa\label{eq:tfds2}
|\Omega_\text{Shell}\rangle &=& \frac{1}{Z^{\frac{1}{2}}}\sum_A\sum_B\lambda_{AB}|A_\text{I}\rangle{\otimes}|B_\text{H}\rangle\,,
\eeqa
where $|A_\text{I}\rangle$ is the outgoing state with $n_{jb}$ particles in the $j$th outgoing mode and 
$|B_\text{H}\rangle$ is the horizon state with $n_{kc}$ in the $k$th mode going into the hole. For more details, 
see~\cite{Hawking,Parker} and~\cite{DeWitt,Fabbri&Navarro-Salas}.

The normalization factor $Z$ allowing to have $\langle\Omega_\text{Shell}|\Omega_\text{Shell}\rangle = 1$, 
is infinite. Indeed, following, for instance,~\cite{Frolov&Novikov,Fabbri&Navarro-Salas}, one can show that
\beqa
Z &=& \text{Tr}\big(\hat{\rho}_{T_\text{H}}\big) \;\;=\;\; \sum_A\sum_B|\lambda_{AB}|^2 \;\;=\;\; \infty\,,
\eeqa
i.e. $\hat{\rho}_{T_\text{H}}$ is not a trace class operator.
This means that the equality~\eqref{eq:tfds2} is merely formal. In other words, one {\sl cannot} 
interpret~\eqref{eq:tfds2} as the initially pure state (the left-hand side of~\eqref{eq:tfds2}) is 
transformed into an entangled state (the right-hand side of~\eqref{eq:tfds2}) during the 
Hawking process. These states are actually orthogonal. If one insists on the equality~\eqref{eq:tfds2}, 
then one violates the unitarity. It is worth emphasizing that this unitarity violation is an origin of the 
ordinary non-unitarity in black-hole physics which was pointed out in~\cite{Hawking}
even if the final state would be still pure. 

This result can be treated as a check of the above statement related to the type III property of the field 
operator algebra $\mathcal{A}(\mathcal{M})$. Thus, one is forced to conclude that 
\beqa
\mathcal{H}_{\Phi} &\not\cong& \mathcal{H}_\text{H}{\otimes}\mathcal{H}_\text{I}\, \quad
\text{or} \quad \mathcal{H}_{\Phi} \;\perp\; \mathcal{H}_\text{H}{\otimes}\mathcal{H}_\text{I}\,.
\eeqa 

\subsection{Proposal}
\label{subsec:proposal}

Our proposal is as follows. One should {\sl not} consider the final scalar particle vacuum state $|0_+\rangle$ as 
realizable physically. 
The reason is that this choice of the final state leads to the unitarity violation by hand.
As discussed above, the physically realizable state is the Unruh vacuum $|\Omega_\text{BH}\rangle$ up to the 
moment of time when, in particular, the backreaction of the quantum field $\hat{\Phi}(x)$ can no longer be 
neglected. Thus, there is {\sl no} evidence of non-unitarity, at least at the level of local quantum field theory.

The black hole singularity potentially invalidates the use of local quantum field theory at the end of the black-hole
evolution. In other words, the subsequent evolution of $|\Omega_\text{BH}\rangle$ or even $|\Omega\rangle$\footnote{This 
would correspond to a phase transition.} 
when $\langle \Omega_\text{BH}|\hat{T}_{\Phi}(x) |\Omega_\text{BH}\rangle \sim \text{O}(1)$
and/or $\langle \Omega_\text{BH}|\hat{T}_\text{m}(x)|\Omega_\text{BH} \rangle \sim \text{O}(1)$ requires either 
quantum gravity or a certain semi-classical 
prescription allowing to describe the complete black-hole disappearance still in the semi-classical language. 
This should be understood in the same spirit of a treatment of the bounce in cosmology by semi-classical models 
(see, e.g. a recent review~\cite{Battefeld&Peter}). 
Therefore, black-hole solutions without curvature singularity~\cite{Klinkhamer} as well
as non-singular ``phenomenological'' black-hole geometries~\cite{Frolov&Vilkovisky,Hayward,Frolov}
are of great interest.

However, as pointed out above, one should not trust in GR in the vicinity of the singularity. Nevertheless,
a decrease of the horizon size is still a reliable prediction. Consequently, the fig. 5 in the second paper 
of~\cite{Hawking1} should \emph{not} be taken for granted. At the level of LQFT, the black-hole evaporation 
might then lead to a disappearance of the horizon inside the support of the black-hole core through properly taking 
into account the backreaction. 

\section{Concluding remarks}
\label{sec:conclusions}

The non-unitarity problem appears, in particular, because the equivalence 
$\mathcal{H}_{\Phi} \cong \mathcal{H}_\text{H}{\otimes}\mathcal{H}_\text{I}$ as well as the equality in~\eqref{eq:tfds2} were 
implicitly assumed. We have shown above that this isomorphism and the equality in~\eqref{eq:tfds2} are actually 
impossible in quantum field theory.

It is also worth emphasizing this as follows. A representation of the Cauchy surface $\Sigma_{\mathcal{M}}$ as
a union $\Sigma_\text{H}{\cup}\Sigma_\text{I}$ does {\sl not} lead to the Hilbert space factorization, i.e.
$\mathcal{H}_{\Phi}$ and $\mathcal{H}_\text{H}{\otimes}\mathcal{H}_\text{I}$ are unitarily inequivalent
in local quantum field theory. However, the contrary is usually employed to formulate the information loss problem 
or the breakdown of \emph{quantum} predictability 
in black-hole physics.

Observer's particle states being elements of $\mathcal{H}_\text{I}$ do not belong to the physical Hilbert space 
$\mathcal{H}$ of the system. That is a thermal gas of the excitations being elements of $\mathcal{H}_\text{I}$ {\sl cannot} be 
represented in the Hilbert space $\mathcal{H}$. The field observables employed by an experimentalist at future time infinity are 
$\hat{\mathcal{O}}_b(x) \in \mathcal{A}(I)$. The non-trivial thermal response of the pure state 
$|\Omega_\text{BH}\rangle$ at future time infinity is due to the fact these operators have vanishing support inside of the black hole. 
Modelling this by a thermal density matrix $\hat{\rho}_{T_\text{H}}$ seems to be misleading, because one then needs to average the field operators 
over the states being {\sl not} represented in the physical Hilbert space, i.e. $\mathcal{H}$. The effect should instead be  understood 
in terms of local probes of system's vacuum $|\Omega_\text{BH}\rangle$ by non-trivial field observables, i.e. $\hat{\mathcal{O}}_b(x)$, by the experimentalist. 

Schematically, an available set of the field observables for the
\emph{stationary} 
experimentalist evolves as follows
\beqa
\hat{\mathcal{O}}_a(x) \;&\; \rightarrow \;&\; \hat{\mathcal{O}}_b(x)\,,
\eeqa
such that system's vacuum $|\Omega_\text{BH}\rangle$ is a Kubo-Martin-Schwinger (KMS)\footnote{The KMS state 
$|\Omega_{\beta}\rangle$ is a state which satisfies the KMS condition: 
$\langle\Omega_{\beta}|\alpha_K^t(\hat{A})\hat{B}|\Omega_{\beta}\rangle = 
\langle\Omega_{\beta}|\hat{B}\alpha_K^{t+i\beta}(\hat{A})|\Omega_{\beta}\rangle$, where both sides are analytic in 
the strip $0 < \text{Im}(t) < \beta$, continuous on its boundary, and 
$\alpha_K^t(\hat{A}) \equiv \exp(+i\hat{K}t)\hat{A}\exp(-i\hat{K}t)$, $\hat{K}$ is a Hermitian operator corresponding to 
the Killing vector $K$. More details can be found, for instance, in~\cite{Haag}.} or thermal state with respect 
to $\partial_{t_\text{S}}$ for operators of the type 
$\hat{\mathcal{O}}_b(x)$ at the Hawking KMS parameter $\beta_\text{H} = 1/T_\text{H}$, and for operators of the type 
$\hat{\mathcal{O}}_a(x)$ at the infinite KMS parameter $\beta$. For instance, a field operator modelling a detector 
considered in~\cite{Fredenhagen&Haag} (see also~\cite{Haag}) belongs to $\mathcal{A}(I)$ and non-trivially responses
when it probes system's vacuum.
This is in complete agreement with the principles of quantum field theory and does not lead 
to the unitarity violation, because system's vacuum $|\Omega_\text{BH}\rangle$ is pure all the time during the black hole formation and its subsequent gradual evaporation. The latter is due to the non-trivial vacuum polarization effect. A representation of the black-hole evaporation as due to the vacuum polarization effect has been suggested in~\cite{Israel1} (see also~\cite{Emelyanov3}). 

It seems the main problem is to take the backreaction of the quantum fields into account 
which should lead to deviations from the perfect thermality of system's vacuum when it is probed by local observables 
available to the \emph{stationary} experimentalist. In other words, the backreaction should lead to
\beqa
\begin{array}{@{}c@{\;}c@{\;}c@{\;}c@{\;}c@{\;}c@{\;}c@{}}
\hat{\mathcal{O}}_a(x) \;\;&\;\; \rightarrow \;\;&\;\; \hat{\mathcal{O}}_b(x) \;\;&\;\; \rightarrow \;\;&\;\; 
\hat{\mathcal{O}}_b(x) + \delta\hat{\mathcal{O}}_b(x) \;\; &\;\;  \not\in \;\; &\;\; \mathcal{A}(I) \\[2mm]
\vin \;\;&&\;\; \vin \;\;&&\;\; \vin \\[2mm]
\mathcal{A}(\mathcal{M}) && \mathcal{A}(I) && \mathcal{A}(\mathcal{M}) && \phantom{1}
\end{array}
\eeqa
such that $|\Omega_\text{BH}\rangle$ is only approximately a KMS state for modified operators 
$\hat{\mathcal{O}}_b(x) + \delta\hat{\mathcal{O}}_b(x)$. It is worth emphasizing that although this would be
a sign of the unitarity restoration, we do not need it, because the process is unitary all the time during the black-hole 
evolution as pointed out above. 

From the argument based on the type III property of the factor operator subalgebras and its various representations, 
we have concluded that a density matrix usually introduced to describe future time experience of an observer is of \emph{no} physical meaning. One may accept the contrary, but then we have a change of the representation of the operator algebra or a phase transition. Consequently, the (Hawking) particles detected at infinity would be physically 
of a different sort in comparison with those from which the collapsed matter shell was composed. For example, this is analogous to the well-known phase transition in QCD, where the change of the Fock space representation leads to a change of the notion of a particle:  
hadrons at low energies and quarks and gluons at energies above the QCD energy scale $\Lambda_\text{QCD} \approx 0.3\;\text{GeV}$.\footnote{In this sense, we have an inequivalence of the Heisenberg and Schr\"{o}dinger pictures in QFT. In certain situations, a quantum system "prefers" to change the representation (e.g. various phase transitions), in others, the algebraic structure (e.g. the Casimir and Hawking effect).} Thus, there are \emph{no} evidences in favour 
of the loss of the quantum coherence working in the framework of the algebraic QFT whenever one does not allow a phase transition during the black-hole formation.\footnote{There is also \emph{no} loss of the quantum coherence
in flat space when the final Cauchy hypersurface fails to be a Cauchy hypersurface for Minkowski
spacetime. This can be seen if one employs the Reeh-Schlieder theorem~\cite{Haag} to the observable
algebra of operators with a support over the final Cauchy hypersurface. Indeed, even with a subalgebra
of the total algebra of all possible field operators, one can probe all elements of the Minkowski Hilbert space 
with an arbitrary precision, whereas the Minkwoski Hilbert space is generated by applying all possible field operators 
from the total algebra. This is not the case if one allows a factorization of the Minkowski Hilbert space
when the final Cauchy hypersurface is merely a subset of the Cauchy hypersurface for Minkowski space.}

Another face of the problem is the non-conservation of baryon number $n_\text{b}$~\cite{Hawking,Wald}. Initially, one 
has $\hat{B}|\Omega_\text{Shell}\rangle = n_\text{b}|\Omega_\text{Shell}\rangle$, 
where $\hat{B}$ is the baryon number operator. 
If the assumption of the 
association of a single particle state with an almost local quantum operator is right, then $\hat{B}$ counts the 
number of these operators building $\hat{\mathcal{O}}_\text{Shell}$. Since the number of these operators is the same 
in $\hat{\mathcal{O}}_\text{BH}$, the baryon number should be conserved, at least till one can no longer neglect the 
backreaction,
because, as pointed out above, the horizon can hide under the black-hole core and the matter can be thrown away
to spacial infinity. 

It would be interesting to investigate our proposal further in detail for a presence of any unphysical or unacceptable 
consequences. 
It is worth noting that we are in agreement with a \emph{general} argument in favour of the unitary 
evaporation of a black hole made in the context of the AdS/CFT conjecture. 

\section*{
ACKNOWLEDGMENTS}

It is a pleasure to thank D. Buchholz, S. Konopka, D. Ponomarev, T. Rug and I. Sachs for discussions. I am especially grateful to M. Haack, 
F. Klinkhamer, A. Vikman for discussions from which I benefited a lot during preparation of this article 
and their valuable suggestions/comments on an early version of this paper. This research is supported by TRR 33
``The Dark Universe''.


\end{document}